%% file: paper.tex
\documentclass[twocolumn]{aastex63}
\usepackage{verbatim}
\usepackage{comment} 

\usepackage{calrsfs}
\DeclareMathAlphabet{\pazocal}{OMS}{zplm}{m}{n}

\makeatletter
\def\@hex@@Hex#1%
 {\if a#1A\else \if b#1B\else \if c#1C\else \if d#1D\else
  \if e#1E\else \if f#1F\else #1\fi\fi\fi\fi\fi\fi \@hex@Hex}
\makeatother
\definecolor{afcolor}{HTML}{b3443c}

\revised{\today}
\shorttitle{Dust in blue monsters}
\shortauthors{A. Ferrara et al.}
\graphicspath{{./}{figures/}}

\begin{document}
\include{definitions}

\title{ALMA observations of super-early galaxies: attenuation-free model predictions}
\correspondingauthor{Andrea Ferrara}
\email{andrea.ferrara@sns.it}
\author[0000-0002-9400-7312]{A. Ferrara}
%\author[0000-0002-7129-5761]{Andrea Pallottini}
\author[0000-0002-6719-380X]{S. Carniani}
\author[0000-0002-9263-7900]{F. di Mascia}
\affil{Scuola Normale Superiore,  Piazza dei Cavalieri 7, 50126 Pisa, Italy}
\author[0000-0002-4989-2471]{R.J. Bouwens}
\affil{Leiden Observatory, Leiden University, NL-2300 RA Leiden, Netherlands}
\author[0000-0001-5851-6649]{P. Oesch}
\affil{Observatoire de Gen{\`e}ve, 1290 Versoix, Switzerland}
\affil{Cosmic Dawn Center (DAWN), Niels Bohr Institute, University of Copenhagen, Jagtvej 128, K\o benhavn N, DK-2200, Denmark}
\author[0000-0001-9746-0924]{S. Schouws}
\affil{Leiden Observatory, Leiden University, NL-2300 RA Leiden, Netherlands}

\begin{abstract}
The abundance and blue color of super-early (redshift $z>10$), luminous galaxies discovered by JWST can be explained if radiation-driven outflows have ejected their dust on kpc-scales. To test this hypothesis, we predict the ALMA detectability of such extended dust component. Given the observed properties of the galaxy, its observed continuum flux at 88 $\mu$m, $F_{88}$, depends on the dust-to-stellar mass ratio, $\xi_d$, and extent of the dust distribution, $r_d$. Once applied to the most distant galaxy known, GS-z14-0 at $z=14.32$, the fiducial model ($\xi_d = 1/529$) predicts $F_{88}^{\rm fid} = 14.9\, \mu$Jy, and a dust extent $r_d=1.4$ kpc. If the galaxy is very dust-rich ($\xi_d =1/40$), $F_{88}^{\rm max} = 40.1\, \mu$Jy. These values are smaller ($F_{88}^{\rm fid} = 9.5\, \mu$Jy) if the dust is predominantly made of large grains as those formed in SN ejecta. Forthcoming ALMA observations might come very close to constraining the fiducial predictions of the outflow-based attenuation-free model. Other super-early galaxies are predicted to be fainter at 88 $\mu$m, mostly because of their lower SFR compared to GS-z14-0, with fiducial fluxes in the range $2-5.2\ \mu$Jy. 
\end{abstract}
\keywords{galaxies: high-redshift, galaxies: evolution, galaxies: formation}

\section{Introduction} \label{sec:Intro}
The approximately fifteen galaxies discovered so far by the \textit{James Webb Space Telescope} (JWST) at redshift $z>10$ 
\citep{Naidu22, Arrabal23, Hsiao23, Wang23, Fujimoto23b, Atek22, Curtis23, Robertson23, Bunker23, Tacchella23, Arrabal23, Finkelstein23, Castellano24, Zavala24, Helton24, Carniani24, Robertson24}
have raised a number of questions concerning early galaxy evolution. Not only their number significantly overshoots extrapolations and models anchored at lower redshifts, but these objects also share some peculiar properties whose origin constitute a challenging puzzle for $\Lambda$CDM-based\footnote{Throughout the paper, we assume a flat Universe with the following cosmological parameters: $\Omega_{\rm M} = 0.3075$, $\Omega_{\Lambda} = 1- \Omega_{\rm M}$, and $\Omega_{\rm b} = 0.0486$,  $h=0.6774$, $\sigma_8=0.826$, where $\Omega_{M}$, $\Omega_{\Lambda}$, and $\Omega_{b}$ are the total matter, vacuum, and baryon densities, in units of the critical density; $h$ is the Hubble constant in units of $100\,\kms$, and $\sigma_8$ is the late-time fluctuation amplitude parameter \citep{planck:2015}.} galaxy formation scenarios. 

At odd with their lower redshift counterparts, many super-early galaxies are characterised by bright UV luminosities ($M_{\rm UV} \simlt -20$), steep UV spectral slopes ($\beta \simlt -2.2$), compact sizes (effective radius $r_e \approx 200$ pc). The analysis of their Spectral Energy Distribution suggests large (for their epoch) stellar masses ($M_\star \approx 10^9 M_\odot$), thus identifying these rare systems as high-$\sigma$ peaks of the cosmic density field populating the bright-end of the luminosity function. Several alternative hypotheses have been suggested to explain the above properties\footnote{These include (a) star  formation variability \citep{Mason23,Mirocha23,Pallottini23}, (b) reduced feedback resulting in a higher star formation efficiency \citep{Dekel23,Li23}, and (c) a top-heavy IMF \citep{Inayoshi22}, although see \citet{Cueto23}. More exotic solutions involving primordial black holes have also been suggested \citep{Liu22}.}. 

According to the so-called \quotes{Attenuation-Free Model} \citep[AFM,][]{Ferrara23a, Ziparo23, Fiore23, Ferrara24a, Ferrara24b} bright luminosities and blue colors result from extremely low dust attenuation conditions. These could be due to low absorbing dust column density, or to grain optical properties producing “gray”, relatively transparent extinction curves \citep[see, e.g.][]{Markov24} at high redshift. 

AFM postulates that dust has been produced in supernova (SN) ejecta with standard net\footnote{That is, after reprocessing by the reverse shock thermalizing the ejecta} yields $y_d \approx 0.1 M_\odot$/SN. After their injection into the interstellar medium, grains could either grow by gas-phase heavy elements accretion and/or be partially destroyed by shocks. However, at $z>10$ the prevailing physical conditions result in timescales  of both processes are longer than the Hubble time, thus making them ineffective \citep{Ferrara16, Hirashita19, Martinez19, Lesniewska19, Dayal22, Priestley21}. The same argument applies to other known dust sources, such as AGB and evolved stars \citep{Ferrarotti06, Valiante09}. 

Corroborated by the evidence for significant amounts of heavy elements in many super-early galaxies \citep{curti2022, Bunker23, Calabro24}, it is physically sound to assume that dust was also copiously produced by the large number ($\approx 10^7$) of SNe in these galaxies. Moreover, due to their very compact size, \citet{Ferrara24a} have shown that in a few Myr the produced dust would completely obscure the galaxy. Instead, \textit{JWST} spectra indicate extremely small V-band attenuations, $A_V \simlt 0.3$. 

To unify several aspects of the problem, it has been proposed that, in spite of the large stellar  masses, attenuation-free conditions can be established by outflows \citep{Ziparo23, Fiore23}. As these objects have most often luminosities exceeding the effective Eddington luminosity for a dusty medium \citep{Ferrara24a}, which can be to first order translated in a lower bound on the specific star formation rate $\rm sSFR > 25\ Gyr^{-1}$, they can launch powerful outflows driven by radiation pressure on the dust itself. 

As $A_V \propto r_d^{-2}$, where $r_d$ is the radial extent of the dust distribution, the outflow can effectively reduce the attenuation by transporting dust on larger spatial scales. For example, expanding the distribution from $r_e\approx 200$ pc to $\approx 2$ kpc, would cause a dramatic drop of the UV optical depth by 100 times, making the galaxy almost transparent and bright.  Thus, AFM predicts that dust should be significantly more extended than the stars in super-early galaxies. Moreover, such dust component should emit thermal far infrared (FIR) continuum due to grain reprocessing of the absorbed UV/optical light from the stars within $r_e$. 

Here we aim at assessing whether the FIR signal produced by such extended dust distribution produced by outflows can be detected by the \textit{Atacama Large Millimeter Array} (ALMA). The detection depends on several factors, such as the dust (i) spatial extent, (ii) mass, and (iii) temperature (including CMB effects), and (iv) grain optical properties. This experiment represents a stringent test for the AFM. 

In this Letter, we mainly concentrate on the specific, but yet utmost important, case of the most distant galaxy known, GS-z14-0 ($z=14.32$, \citealt{Carniani24}) for which the required ALMA data will become soon available (DDT 2023.A.00037.S, PI: S. Schouws).  

A positive detection would strongly support the idea that dust is produced by SNe, and subsequently ejected by outflows outside the galaxy main body.  A non detection would instead put meaningful upper limits on early dust production and properties at these yet basically unexplored cosmic epochs. We also apply the model to four additional $z>10$ galaxies.

%
% FIGURE 1
%
\begin{figure*}
\centering\includegraphics[width = 1.0 \textwidth]{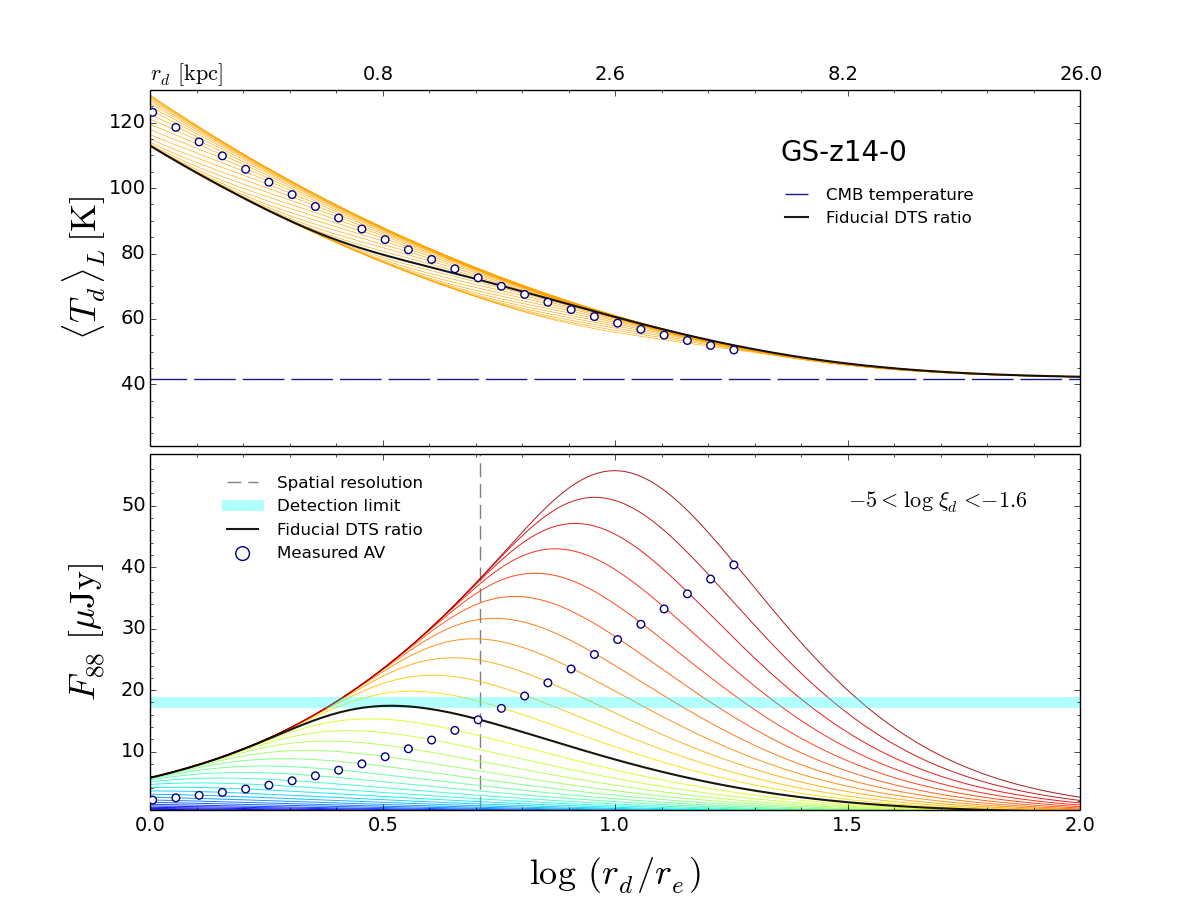}
\caption{\textit{Top panel}: Temperature distribution as a function of the dust spatial distribution radius, $r_d$, in units of the effective stellar radius of GS-z14-0, $r_e=260$ pc. The different orange curves span the dust-to-stellar ratio range $-5 < \log \xi_d < -1.6$ equally spaced in 0.1 logarithmic intervals, with the black curve highlighting the fiducial case $\log \xi_d = -2.72 (\xi_d = 1/529)$. Also shown is the CMB temperature at the galaxy redshift $z=14.32$, The circles indicate the radius for which $A_V$ of the galaxy matches the observed one, $A_V = 0.31$. \textit{Bottom panel}: As in the top panel, but for the predicted observed flux from GS-z14-0 at 88$\mu$.  Also shown are the expected 3$\sigma$ sensitivity limit ($18 \mu$Jy/beam), and spatial resolution (0.418”) of the DDT 2023.A.00037.S (PI: S. Schouws) observations.
}
\label{fig:Fig01}
\end{figure*}
\section{FIR dust continuum emission}\label{sec:FIRemission}
To predict the expected FIR continuum emission, we first quantify the dust production and its spatial distribution. Next, we compute the dust temperature including CMB and radiative transfer effects. From these quantities we finally obtain the observed FIR flux from a given galaxy.

\subsection{Dust production}
At high redshift ($z>7$) dust production can be safely assumed to be largely dominated  by SNe \citep{Todini01, Lesniewska19, Dayal22, Sommovigo22, Ferrara22a,Schneider24} as grain growth \citep{Ferrara16, Hirashita19} is sub-dominant, and other known sources (AGB and evolved stars, \citealt{Valiante09}) have evolutionary times  exceeding the Hubble time (294 Myr at $z=14.32$). 

We assume that stars form according to a Salpeter Initial Mass Function in the range $1-100\ M_\odot$. As a result, a SN occurs every $\nu^{-1} = 52.9 M_\odot$ of stars formed. 

The dust yield per SN is rather uncertain as it involves complex nucleation physics of heavy elements in the SN ejecta. Also, newly formed grains are subject to destruction processes when they pass through the reverse shock. The net production yield is therefore poorly constrained, but based also on local SN studies \citep{Matsuura15, Rho18, Priestley21, Niculescu21}, it ranges from 0.01 to 1.1 $M_\odot$. Therefore we use $y_d = 0.1 \ M_\odot$ as a fiducial value. 

Given the IMF, this corresponds to a dust-to-stellar ratio $\xi_d = y_d\nu = 1/529$. However, given the above uncertainties, we explore the implications of the wide range of values $-5  < \log \xi_d < -1.6$. This range should therefore encompass both a situation in which freshly formed dust is heavily destroyed by the reverse shock, and a super-efficient dust condensation under very favourable conditions. 

For the fiducial value of $\xi_d$ and the measured stellar mass of GS-z14-0 ($M_\star = 4\times 10^8 M_\odot$; \citealt{Carniani24}), we find that $7.5\times 10^5 M_\odot$ of dust should have been produced by the time of the observation. {We recall that dust typically condenses in the ejecta few hundreds of days after the explosion \citep{Todini01, Nozawa07}. For our purposes dust production can be considered as instantaneous, and occurring on the same evolutionary timescale of SN massive star progenitors, i.e., $\simlt 20$ Myr}.

\subsection{Dust spatial distribution}
Initially, the dust is distributed as the observed massive stars providing the observed rest frame UV luminosity. In GS-z14-0, these are distributed within a measured effective radius $r_e=260$ pc. Under this hypothesis, the dust optical depth at 1500~\AA\ would be equal to $\tau^e_{1500} = (\kappa_{1500} \xi_d/4 \pi r_e^2)M_\star = 23.3$, having used a value of the dust mass absorption coefficient  $\kappa_{1500} = 1.26\times 10^5\ \rm cm^2\ g^{-1}$ appropriate for a MW-like extinction curve \citep[][WD01]{Weingartner01}. 

By translating the observed $A_V = 0.31$ value into an optical depth, using $\tau_{1500} = 2.655(A_V/1.086)$, where the pre-factor accounts for the differential attenuation between 1500 \AA\ and the V-band for a MW curve, we find $\tau_{1500} = 0.76$, a value $\approx 30$ times smaller than $\tau^e_{1500}$.  

Using the AFM to predict the star formation history of GS-z14-0, \citet{Ferrara24b} found that this galaxy has recently undergone a super-Eddington phase in which a powerful outflow has mini-quenched its star formation leaving it in the observed post-stardust phase. The outflow has ejected a significant fraction of the dust, metals and gas previously contained in $r_e$. 

From the previous estimates we see that to decrease $A_V$ to its observed value, the dust has to be re-configured by the outflow into a much more extended distribution of radius $r_d = (\tau^e_{1500}/\tau_{1500})^{1/2} r_e = 5.5 r_e$, which corresponds to $r_d \simeq 1.4\ \rm kpc$, for the fiducial $\xi_d$ value. 

Without the outflow hypothesis, to make the galaxy optically thin one should assume an extremely low dust content, corresponding to $\xi_d=6.6\times 10^{-5}$. For comparison, for galaxies in the REBELS sample \citep{Bouwens22a}  $\xi_d \approx 0.01$ \citep{Ferrara22a, Dayal22, Sommovigo22}. The results presented below explore both (and additional) possibilities showing that ALMA observations can clearly discriminate between these two alternative scenarios by detecting the FIR dust emission.  

%
% FIGURE 2
%
\begin{figure*}
\centering\includegraphics[width = 1.0 \textwidth]{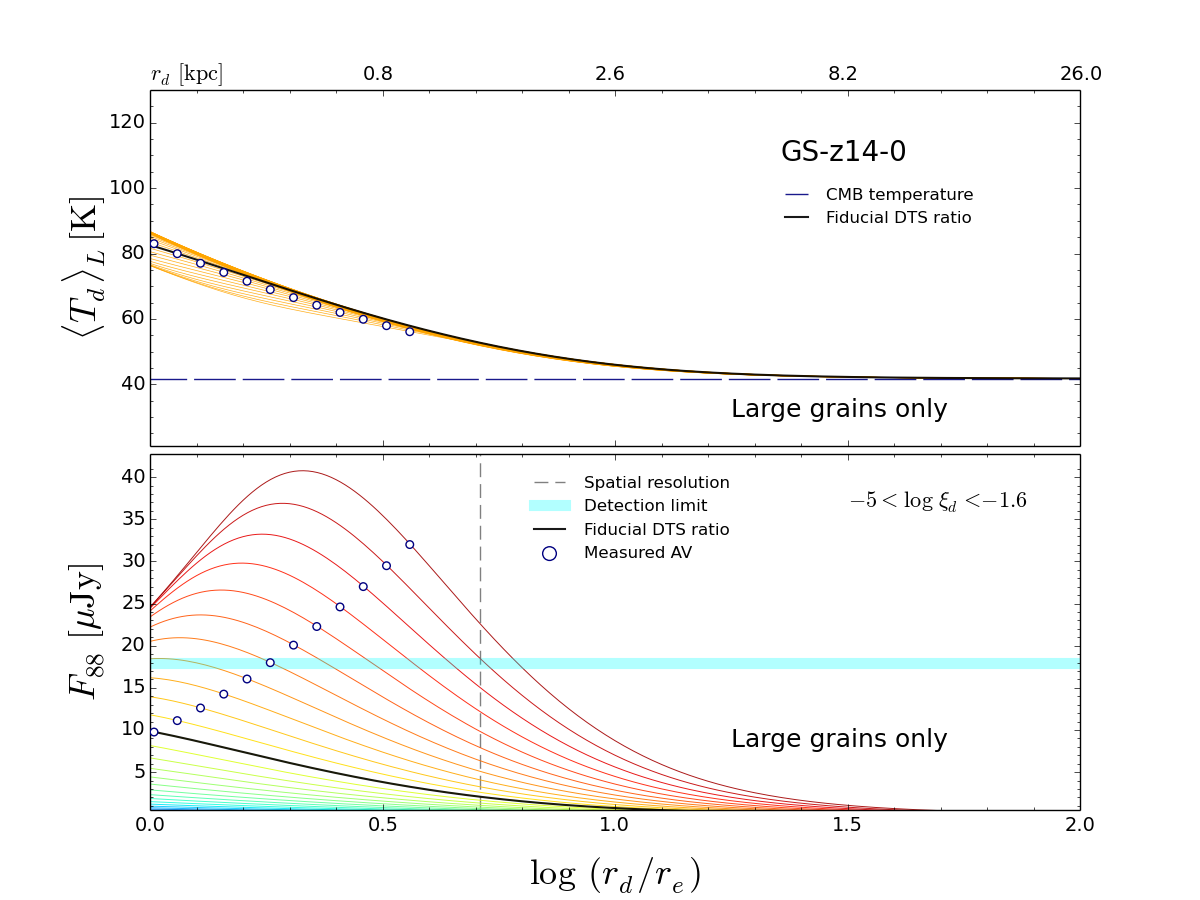}
\caption{Same as Fig. \ref{fig:Fig01}, but assuming that the dust is constituted by large silicate grains with a log-normal distribution centered at 0.5 $\mu$m, appropriate for grains freshly formed in SN ejecta.
} 
\label{fig:Fig02}
\end{figure*}

\subsection{Dust temperature}
Dust grains are heated by absorption of UV photons whose energy is re-emitted in the FIR. Depending on the assumed spatial distribution, grains can achieve different temperatures, thus affecting the emitted FIR flux. The latter is computed here at a rest frame  wavelength of 88$\mu$m since high-$z$ ALMA observations are often tuned to detect [OIII] 88$\mum$ line emission \citep{Popping22, Bakx22, Kaasinen22, Fujimoto23c, Kohandel23} 

The emitted radiation spectrum is classically modelled \citep[e.g.][]{Dayal10, Hirashita14} as a grey-body at mean dust temperature, $\bar T_d$, given by 
\begin{equation}\label{eq:Tmpt}
    \bar T_d = \left(\frac{L_{\rm abs}}{\Theta M_d}\right)^{1/(4+\beta_d)},
\end{equation}
where
\begin{equation}
    \Theta = \frac{8\pi}{c^2}\frac{\kappa_{88}}{\nu_{88}^{\beta_d}}\frac{k_B^{4+\beta_d}}{h_P^{3+\beta_d}}\zeta(4+\beta_d)\Gamma(4+\beta_d),
\end{equation}
and $L_{\rm abs} = L_{\rm bol} (1-e^{-\tau_{1500}})$ is the fraction of the stellar UV radiation that is absorbed by dust. The galaxy bolometric luminosity is written as $L_{\rm bol} = f_{\rm bol} L_{1500}$, where the intrinsic galaxy luminosity at 1500\AA\ ($L_{1500}$) is obtained from the inferred star formation rate (${\rm SFR}$), via a conversion factor, ${\cal K}_{1500}$ $[{L_\odot}/M_\odot {\rm yr}^{-1}]$, whose value has been chosen so to match the one used by the ALMA REBELS survey \citep{Bouwens22a}: ${\cal K}_{1500} \equiv {L_{1500}}/{\rm SFR} = 0.587 \times 10^{10}$. Following \citet{Fiore23}, we adopt $f_{\rm bol} = 2$.

We take the mass absorption coefficient $\kappa_{88}$ and $\beta_d$ consistently with the adopted WD01 extinction curve 
$\kappa_{88}= 34.15\, {\rm cm}^2 {\rm g}^{-1}$, and $\beta_d=2.03$; $\zeta$ and $\Gamma$ are the Zeta and Gamma functions, respectively; finally, $\nu_{88}$ is the frequency corresponding to wavelength $\lambda =88\ \mu$m. Other symbols have the usual meaning. %We then find $\Theta = (4.89, 5.33)\times 10^{-6}$ for (MW, SMC). 

The temperature in eq. \ref{eq:Tmpt} is a mean (or \quotes{effective}) physical dust temperature such that energy balance between dust absorbed and emitted photon energy is guaranteed. In general, though, radiative transfer effects produce a temperature distribution, with $T_d$ decreasing away from the source. It is therefore necessary to include a correction factor which becomes important when the system becomes optically thick. 
%
%Eq. \ref{eq:Tmpt} holds precisely if the dust is piled-up by the outflow (as expected) in a geometrically thin shell. If instead dust is uniformly distributed in a sphere of the same radius, one can show that $T_d$ is not constant even in the optically thin case, and $T_d \propto r^{-1/3}$. If the shell is optically thick a $T_d$ gradient will also develop in the radial direction. To avoid these complications, we assume a shell geometry for which it is safe to use eq. \ref{eq:Tmpt} in the optically thin case, and apply a luminosity-weighted correction for the optically thick one.
%
%Such value corresponds to the temperature dust grains would attain should the available UV energy being uniformly distributed among them. This is possible only if the system is optically thin, i.e. $\tau_{1500} \ll 1$. In general, though, radiative transfer effects produce a temperature distribution, with $T_d$ decreasing away from the source. 
%

\citet[][see their Appendix A]{Ferrara22a} showed that the luminosity-weighted temperature, $\langle T_d \rangle_L$, of an absorbing dust slab depends on its total optical depth, and can be written as
\be\label{eq:TL}
\langle T_d \rangle_L = \bar T_d\, \frac{6}{7}\tau_{1500}^{1/b}\frac{(1 - e^{-7\tau_{1500}/b})}{(1 - e^{-\tau_{1500}})^{7/b}}  \equiv \bar T_d f_L(\tau_{1500}),
\ee
where $b = 4+\beta_d$. Applying eq. \ref{eq:TL} results 
%\footnote{Energy conservation requires that the total $M_d$ in eq. \ref{eq:Tmpt} must be reduced by a factor $f_L^{-(4+\beta_d)}$, which might result in a $(2-3)\times$ lower mass estimate. In the following we denote this reduced dust mass by $M_d' = f_L^{{-(4+\beta_d)}} M_d $.} 
in temperatures that are generally higher\footnote{We neglect self-heating by the dust, i.e. absorption of thermal IR radiation emitted by the grains, as it becomes important at high optical depth \citep{Gordon17}, while the configurations of interest here have $A_V <1$. } than $\bar T_d$. To conserve energy $M_d$ must be reduced by a factor $f_L^{-(4+\beta_d)}$. In the following we denote this reduced dust mass by $M_d' = f_L^{{-(4+\beta_d)}} M_d $.

% ______________________________________________________
\subsection{Observed flux at  88$\mu$m}
From the previous results it is straightforward to compute the restframe 88 $\mu$m specific flux from a source at $z=z_s$ observed at wavelength $88(1+z_s)\, \mu$m:
\begin{equation}\label{eq:F88}
    F_{88} = g(z_s)\, \kappa_{88} M'_d [B_{88}(T'_d)-B_{88}(T_{\rm CMB})];
\end{equation}
$B_\lambda$ is the black-body spectrum, and $T_{\rm CMB}(z)=T_0(1+z_s)$, with $T_0=2.7255$ K  \citep{Fixsen09} is the CMB temperature at redshift $z_s$. Equation \ref{eq:F88} accounts for the fact that the CMB acts as a thermal bath for dust grains, setting a lower limit to their temperature. At $z_s=14.32$ such minimum temperature corresponds to $T_{\rm CMB}=41.75$ K. Finally, $T'_d$ is the CMB-corrected dust temperature\footnote{In the remainder of the paper we will always refer to dust temperature as the \textit{CMB-corrected} one, i.e. $T'_d$ in eq. \ref{eq:CMB}.} following \cite{daCunha13}, 
\begin{equation}\label{eq:CMB}
T'_d = \{\langle T_d \rangle_L^{4+\beta_d}+T_0^{4+\beta_d}[(1+z)^{4+\beta_d}-1]\}^{1/(4+\beta_d)}.
\end{equation}

\section{Results}\label{sec:Results}
Fig. 1 summarises our predictions for the ALMA dust continuum emission from GS-z14-0. The top (bottom) panel shows the dependence of the luminosity-weighted dust temperature (88 $\mu$m flux) for a spherical dust distribution of varying radius $r_d$, normalised to the observed effective stellar radius, $r_e$, of the galaxy. Each curve corresponds to a dust-to-stellar ratio, $\xi_d$, in the range $-5 < \log \xi_d < -1.6$ equally spaced in 0.1 logarithmic intervals with the black curve highlighting the fiducial case $\log \xi_d = -2.72$, or $\xi_d = 1/529$.

The luminosity-weighted temperature trend can be understood with the help of eq. \ref{eq:TL}. Along each orange curve the dust mass $M_d = \xi_d M_\star$ is fixed. The most evident feature is the dust temperature decrease from $110-125$ K at $r_d=r_e$ to the CMB temperature at the largest considered radii ($17.5$ kpc), where the dust becomes virtually undetectable. The temperature drop is due to the fact that as we consider more extended, and therefore more optically thin, dust distributions, the absorbed stellar radiation decreases, and the dust becomes colder. Note that if the dust had not been ejected from the galaxy, and therefore $r_d=r_e$, the dust-to-stellar ratio should be very small ($\log \xi_d = -4.2$) to yield the measured $A_V=0.31$ (denoted by circles).

On top of this trend, at any given radius the curves differ by up to $10-20$\% as a result of variations of $M_d$ from curve to curve, and due to the fact that for small radii, most curves (specifically, those with $\log \xi_d > -4.2$) correspond to optically thick regimes for which $\langle T_d\rangle_L$ is slightly larger (see eq. \ref{eq:TL}). This is easily seen by taking the optically thick limit $\tau_{1500} \to \infty$, in which $\langle T_d \rangle_L \propto\tau_{1500}^{1/(4+\beta_d)}$. For each curve the circles mark the radius at which the attenuation matches the observed value $A_V=0.31$ (higher/lower $A_V$ values to the left/right). For each value of $\xi_d$, the temperature indicated by the circle is then the \quotes{true} (i.e. predicted) one, within model assumptions.

%
% TABLE
%
\begin{table*}
%\begin{minipage}{170mm}
\begin{center}
\caption{Relevant properties of the selected super-early galaxies at $z>10$}
\begin{tabular}{lcccccccc}
\hline\hline
ID             & redshift&      $A_V$ [mag]      & $\rm SFR\ [\msunyr]$  &  $\log(M_\star/\msun)$ &$r_e$ [pc]       & $F_{88}^{\rm fid} [\mu$Jy] & $F_{88}^{\rm max} [\mu$Jy]& $\log(r_d^{\rm fid}/r_e)$ \\
\hline 
GS-z14-0$^a$   & 14.32   & $0.31^{+0.14}_{-0.17}$& $22^{+6}_{-6}$        & $8.60^{+0.7}_{-0.2}$ &$260^{+2}_{-2}$    &          14.9              &  40.1                & 0.73 \\
GS-z14-1$^a$   & 13.90   & $0.20^{+0.11}_{-0.07}$& $2^{+0.7}_{-0.4}$     & $8.00^{+0.4}_{-0.4}$ &$< 160$            &           2.0              &  3.8                 & 0.70 \\
UNCOVER-z12$^b$& 12.39   & $0.19^{+0.17}_{-0.10}$& $3.08^{+1.81}_{-0.68}$& $8.35^{+0.21}_{-0.16}$ &$426^{+40}_{-42}$&           4.4              &  8.5                 & 0.47 \\ 
GS-z11-0$^c$   & 11.58   & $0.18^{+0.06}_{-0.06}$& $2.20^{+0.28}_{-0.22}$& $8.67^{+0.08}_{-0.13}$ &$77^\dag$        &           5.2              &  7.8                 & 1.38 \\
GHZ2$^d$       & 12.34   & $0.04^{+0.07}_{-0.03}$& $5.2^{+1.1}_{-0.6}$   & $9.05^{+0.10}_{-0.25}$ &$105^{+9}_{-9}$  &           3.4              &  4.0                 & 1.76 \\
\hline
%
% Refs: Harikane23, Finkelstein22 [for r_e] (CEERS2); Hsiao+23 (MACS0647-JD); Curtis-Lake23, Robertson+23 [for r_e], Carniani+24 (GS-zX-Y); Bunker (GN-z11); Arrabal-Haro+23, Finkelstein+22 ApJL940 [for r_e] (Maisie); Castellano24 (GHZ2); Wang+23 ApJL957, Fujimoto+23arXiv:2308.11609v3, Atek23MNRAS 524, 5486 (UNCOVER; Note UNCOVER37126 is UNCOVER39074 in Atek23).
\label{tab:properties}
\end{tabular}
\end{center}
%\end{minipage}
\tablecomments{The measured values are taken from the following references: 
$^a$\citet{Carniani24, Robertson24}, $^b$\citet{Wang23, Fujimoto23b, Atek22}, $^d$\citet{Curtis23, Robertson23}, $^d$\cite{Castellano24}; $^\dag$Error not provided. The last three columns show the predictions of the model for the fiducial (maximum) 88$\mu$m observed flux, and the fiducial dust-to-stellar radius ratio for which $A_V$ matches the observed value, respectively.} 
\end{table*}

The behavior of $F_{88}$ in the optically thick regime can be understood as follows. From eq. \ref{eq:F88}, further neglecting CMB effects -- which is appropriate given that dust is hot at small radii, it follows that $F_{88} \propto M' B_\nu(\langle T_d \rangle_L) \simeq M' \langle T_d \rangle_L$ in the Rayleigh-Jeans regime. We have already seen that $\langle T_d \rangle_L \propto \tau_{1500}^{1/(4+\beta_d)}$; also, $M_d' = f_L^{{-(4+\beta_d)}} M_d \propto \tau_{1500}^{-1}$. Hence, $F_{88}$ decreases as $\tau_{1500}^{-(3+\beta_d)/(4+\beta_d)}$, which is exactly the trend shown in Fig. \ref{fig:Fig01}.

The corresponding $88 \mu$m flux is shown in the bottom panel, where it is also compared with the expected 3$\sigma$ continuum sensitivity ($18\ \mu$Jy/beam) and spatial resolution (0.451") of the ALMA DDT 2023.A.00037.S (PI: S. Schouws) observations of GS-z14-0. For each $\xi_d$ value, the flux raises, reaches a peak (approximately where $\tau_{1500}\approx 1)$, and decreases at large radii. The amplitude of the peak correlates with $\xi_d$. As the distribution becomes optically thick for small $r_d/r_e$ values, the curves with larger $\xi_d$ (or, equivalently, larger $M_d$) converge to a single line.

By looking at the circles ($A_V=0.31$) we see that the fiducial model (black line, $\xi_d = 1/529$) predicts a 88$\mu$m flux $F_{88}^{\rm fid}=14.9\ \mu$Jy, essentially at the 3$\sigma$ detection limit of the experiment, and a luminosity-weighted temperature of 72 K. As $\xi_d$ is decreased, implying an efficient grain destruction by interstellar shocks, the signal becomes weaker and more difficult to detect. However, if the galaxy is dust-rich, at or above the level observed in some of the REBELS galaxies showing $\xi_d \approx 0.01$ \citep{Ferrara22a, Dayal22, Inami22}, the signal becomes detectable. 

It is important to recall that the observed $F_{88}$ measurement (or upper limit) should be interpreted along with the constraints provided by the observed $A_V$ value. For example, suppose that -- consistently with the fiducial model in Fig. \ref{fig:Fig02} -- $F_{88}$ is measured to be $14.9\ \mu$Jy (circle on black curve). Then, we would conclude that $\xi_d = 1/529$, and $\log(r_d^{\rm fid}/r_e) \simeq 0.7$, or $r_d=1.3$ kpc. If correct, ALMA observations should also show that the dust is spatially resolved, and more extended than the stellar component. The other, more compact solution for the same $F_{88}$, i.e. $\log(r_d/r_e) \simeq 0.4$, should be discarded because the galaxy would have an $A_V$ much larger than observed.  

A natural case to explore is the one in which the dust, in the absence of any ejective event, remains within $r_e$. In this scenario, the temperature would be significantly higher ($\simeq 100$ K) and the distribution optically thick. These two properties act as to dramatically suppress the flux: independently of $\xi_d$ (unless extremely small), $F_{88} = 5.3\ \mu$Jy. Detecting such a low flux would require very long integration times.  

In conclusion, a positive detection would strongly support the hypothesis that dust produced by SNe associated with the observed stars in GS-z14-0 has been carried to larger galactocentric radii by outflows, making the galaxy bright and transparent to UV radiation. %
%If instead the dust would have remained within $r_e$, we predict that it should not be detected by the ALMA DDT (and hardly from future similar observations), mainly because too hot. 
This conclusion highlights the key importance of ALMA deep observations of super-early galaxies.

\subsection{Large grains}
In addition to dust ejection or destruction by shocks, we need to consider a third possibility, i.e., that early dust is constituted by large, $0.1-1\ \mu$m size grains. The need is based on two facts: (a) theoretically, it is predicted that SN dust lacks small grains which are preferentially destroyed in the ejecta by the reverse shock \citep{Nozawa07, Asano13, Hirashita19}, (b) experimentally, the SED of galaxies in $z=2-12$ \textit{JWST}-detected galaxies show an increasingly flatter extinction curve towards high-$z$ \citep{diMascia21b, Markov24}. These two evidences are clearly connected as large grains of size $a$ absorb light with wavelength $\lambda \ll a$ in a frequency-independent way. In this case, the absorption cross section should be equal to the geometric one, and the UV absorption coefficient $\kappa_{\rm UV}^{\rm large} \simeq 3/4 \delta_g a$, where $\delta_g = 2.95\ \rm g cm^{-3}$ is the material density of silicate grains, i.e. those preferentially produced by SNe. Note that if $a = 0.5\ \mu$m, $\kappa_{\rm UV}^{\rm large} = 0.04 \kappa_{\rm UV}$. 

These estimates are confirmed by detailed calculations performed by \citet{Ysard19} that we use here. These authors compute the optical properties of $0.01\ \mu$m to 10 cm grains from effective medium and Mie theories. To this aim they assume a lognormal distribution of grain sizes centered at a given radius, $a_0$, which we assume to be equal to 0.5 $\mu$m according to the previous discussion, and a fixed dispersion $\sigma = 0.7$. With these assumptions they indeed find  $\kappa_{\rm UV}^{\rm large} \simeq 3/4 \delta_g a=5085\ \rm cm^2 g^{-1}$, and $\kappa_{88}^{\rm large} = 15\ \rm cm^2 g^{-1}$.      

Fig. \ref{fig:Fig02} presents the temperature and expected flux if dust is composed by large $\simeq 0.5\ \mu$m grains. The temperature follows the same trend as in the standard case, but it is significantly colder, reaching at $r_e$ values in the range $75-85$ K. As a result, for any given $\xi_d$, the flux is smaller, e.g. $F_{88}^{\rm fid} = 9.5\ \mu$Jy, and shifted to smaller radii compared to the standard case. Also, the lower opacity of large grains allows the galaxy to become optically thin (as observed) for more compact distributions.

\subsection{Other super-early galaxies}
Our model can be used to make predictions on the FIR observability of other potentially detectable super-early galaxies. In addition to GS-z14-0, which has the largest $A_V$ among $z>10$ spectroscopically confirmed galaxies, we have selected the next three dustiest galaxies, GS-z14-1 ($A_V=0.20$), UNCOVER-z12 (0.19), GS-z11-0 (0.18). We have also included GHZ2  in spite of its very low attenuation ($A_V=0.04$, \citealt{Castellano24}, but see \citealt{Zavala24} who find $A_V=0.01-0.2$ depending of the SED fitter used) because of its very interesting spectrum showing many prominent emission lines. 

The relevant properties (redshift, $A_V$, SFR, $M_\star$, $r_e$) of the galaxies in the sample are reported in Tab. \ref{tab:properties}, along with the predicted value of the fiducial 88$\mu$m flux, and the radius $r_d^{\rm fid}$ where the attenuation becomes equal to the observed $A_V$ value. Also shown is the maximum expected flux, $F_{88}^{\rm max}$ which we take from the curve with the maximum assumed value of $\log\xi_d=-1.6$.  

For the galaxies in the sample the dust temperature and the FIR flux follow very similar trends with $r_d/r_e$ as shown in Fig. \ref{fig:Fig01} for GS-z14-0, albeit they quantitatively differ as a result of the different properties of the galaxy. From the Table, we see that the expected fiducial fluxes range from $2\ \mu$Jy for the faintest galaxy (GS-z14-1) to 14.9 $\mu$Jy for the brightest (and most distant) one, GS-z14-0; the latter therefore remains the most promising target to constrain the dust content of the very early galaxies.
For GHZ2, if we instead assume the largest attenuation found by \citet{Zavala24}, $A_V=0.2$, we find a larger fiducial and maximum flux, $F_{88}^{\rm fid}= 11.2\ \mu$Jy, and $F_{88}^{\rm max}= 16.3\ \mu$Jy, respectively.

Finally, notice that we did not considered the remarkable galaxy GN-z11 \citep{Bunker23} as it is not visible by ALMA due to its northern emisphere location. However, this galaxy has been observed by the Northern Extended Millimeter Array (NOEMA, \citealt{Fudamoto23}). These observations, have put a 1$\sigma$ upper limit of 13 $\mu$Jy/beam at 160$\mu$m. When applied to GN-z11, our model (fiducial case) predicts a maximum flux $F_{160}^{\rm max} = 14.1\ \mu$Jy. Although interesting, available NOEMA observations are unfortunately too shallow to detect the expected signal.    

%\subsection{ALMA observational requirements}

\section{Summary}
To test one of the predictions of the \quotes{Attenuation-Free Model} \citep{Ferrara23a, Ferrara24a}, we have modelled the FIR dust continuum emission from the extended dust distribution produced by outflows around \quotes{blue monsters}, i.e. luminous, blue, super-early ($z>10$) galaxies. Given the observed properties of the galaxies (redshift, $A_V$, SFR, $M_\star$, $r_e$), the expected flux at 88 $\mu$m depends on the dust-to-stellar ratio, $\xi_d$, and extent of the dust distribution, $r_d$. A signal detection with ALMA would therefore constrain the amount of dust and its extent, thus representing a successful test that the attenuation-free conditions are produced by dust ejection by (likely, radiation-driven) outflows. We have discussed in detail GS-z14-0 ($z=14.32$), and further applied the model to other 4 super-early galaxies. We find that:

\begin{itemize}
\item[{\color{red} $\blacksquare$}] If the dust is in the galaxy ($r_d=r_e$) its temperature is $110-125$ K. In this case, though, the dust-to stellar mass ratio must be very small ($\log \xi_d < -4.2$) not to exceed the measured $A_V=0.31$. For $r_d > r_e$, $\langle T_d \rangle_L$ decreases asymptotically towards $T_{\rm CMB}$ (Fig. \ref{fig:Fig01}, top panel).

\item[{\color{red} $\blacksquare$}] The fiducial model ($\xi_d = 1/529$) predicts for GS-z14-0 an observed flux $F_{88}^{\rm fid} = 14.9\ \mu$Jy, and a dust extent $r_d \sim 1.4$ kpc (Fig. \ref{fig:Fig01}, bottom panel). For the largest value considered, $\log \xi_d =-1.6$, $F_{88}^{\rm max} = 40.1\ \mu$Jy. These values are smaller ($F_{88}^{\rm fid} = 9.5\, \mu$Jy) if the dust is predominantly made of large grains as those formed in SN ejecta (Fig. \ref{fig:Fig02}). 

\item[{\color{red} $\blacksquare$}] Forthcoming ALMA observations, with an expected 3$\sigma$ sensitivity of $18 \mu$Jy, should come very close to constraining the fiducial predictions of the outflow-based attenuation-free model. 

\item[{\color{red} $\blacksquare$}] Other super-early galaxies are predicted (Tab. \ref{tab:properties}) to be fainter at 88 $\mu$m, mostly because of their lower SFR compared to GS-z14-0 but also due to their lower $A_V$, with fiducial fluxes in the range $F_{88}^{\rm fid} = 2-5.2\ \mu$Jy. 
\end{itemize}

We conclude by warning that albeit our model is built to catch the main features of the AFM, its simple incarnation cannot include potentially non-negligible effects that might quantitatively change our conclusions. For example, the present study does not account for self-heating of the dust by re-processed thermal photons, dust scattering and grain size distribution. The impact of these effects on the results presented here can only be ascertained via dedicated numerical radiative transfer simulations.

%\section*{Data Availability}
%Data available on request.

\acknowledgments
We thank J. Zavala and M. Castellano for useful discussions, data and comments. 
This work is supported by the ERC Advanced Grant INTERSTELLAR H2020/740120, and in part by grant NSF PHY-2309135 to the Kavli Institute for Theoretical Physics (KITP). 
Plots in this paper produced with the \textsc{matplotlib} \citep{Hunter07} package for \textsc{PYTHON}.    

\bibliographystyle{aasjournal}
\bibliography{paper}

\end{document}

%% file: definitions.tex
%%%%% AUTHORS - PLACE YOUR OWN COMMANDS HERE %%%%%

% Please keep new commands to a minimum, and use \newcommand not \def to avoid
% overwriting existing commands. Example:
%\newcommand{\pcm}{\,cm$^{-2}$}	% per cm-squared

% quick alias
\def\be{\begin{equation}}
\def\ee{\end{equation}}
\newcommand\code[1]{\textsc{\MakeLowercase{#1}}}
\newcommand\quotesingle[1]{`{#1}'}
\newcommand\quotes[1]{``{#1}"}
\def\gsim{\lower.5ex\hbox{\gtsima}} 
\def\lsim{\lower.5ex\hbox{\ltsima}} 
\def\gtsima{$\; \buildrel > \over \sim \;$} 
\def\ltsima{$\; \buildrel < \over \sim \;$} \def\gsim{\lower.5ex\hbox{\gtsima}} 
\def\lsim{\lower.5ex\hbox{\ltsima}} 
\def\simgt{\lower.5ex\hbox{\gtsima}} 
\def\simlt{\lower.5ex\hbox{\ltsima}}

% solar stuff units
\def\msun{{\rm M}_{\odot}}
\def\lsun{{\rm L}_{\odot}}
\def\dsun{{\cal D}_{\odot}}
\def\fsun{\xi_{\odot}}
\def\zsun{{\rm Z}_{\odot}}
\def\msunyr{\msun {\rm yr}^{-1}}
\def\gdens{\msun\,{\rm kpc}^{-2}}
\def\sfrdens{\msun\,{\rm yr}^{-1}\,{\rm kpc}^{-2}}

% units
\def\mum{\mu {\rm m}}
\newcommand{\angstrom}{\mbox{\normalfont\AA}}
\def\cc{\rm cm^{-3}}
\def\uflux{{\rm erg}\,{\rm s}^{-1} {\rm cm}^{-2} }

% quantities defs
\def\fdust{\xi_{d}}
\def\fesc{f_{\rm esc}\,}
\def\td{\tau_{sd}}
\def\Sg{$\Sigma_{g}$}
\def\S*{$\Sigma_{\rm SFR}$}
\def\Ssfr{\Sigma_{\rm SFR}}
\def\Sgas{\Sigma_{\rm g}}
\def\Sstar{\Sigma_{\rm *}}
\def\Sesc{\Sigma_{\rm esc}}
\def\Srad{\Sigma_{\rm rad}}

% additional wrappers for quantities and units
\def\Dsolar{${\cal D}/\dsun$}
\def\Zsolar{$Z/\zsun$}
\def\DDsolar{\left( {{\cal D}\over \dsun} \right)}
\def\ZZsolar{\left( {Z \over \zsun} \right)}
\def\kms{{\rm km\,s}^{-1}\,}
\def\skms{$\sigma_{\rm kms}\,$}

% LINES
\def\Scii{$\Sigma_{\rm [CII]}$}
\def\Sciimax{$\Sigma_{\rm [CII]}^{\rm max}$}
\def\CII{\hbox{[C~$\scriptstyle\rm II $]~}}
\def\CIII{\hbox{C~$\scriptstyle\rm III $]~}}
\def\OII{\hbox{[O~$\scriptstyle\rm II $]~}}
\def\OIII{\hbox{[O~$\scriptstyle\rm III $]~}}
% IONS
\def\HH{\hbox{H$_2$}~} 
\def\HI{\hbox{H~$\scriptstyle\rm I\ $}} 
\def\HII{\hbox{H~$\scriptstyle\rm II\ $}} 
\def\CIion{\hbox{C~$\scriptstyle\rm I $~}}
\def\CIIion{\hbox{C~$\scriptstyle\rm II $~}}
\def\CIIIion{\hbox{C~$\scriptstyle\rm III $~}}
\def\CIVion{\hbox{C~$\scriptstyle\rm IV $~}}
% ion variables
\def\nhh{n_{\rm H2}}
\def\nhi{n_{\rm HI}}
\def\nhii{n_{\rm HII}}
\def\fhh{x_{\rm H2}}
\def\fhi{x_{\rm HI}}
\def\fhii{x_{\rm HII}}
% MIX
\def\fd{f^*_{\rm diss}} 
\def\ks{\kappa_{\rm s}}

% COLORS
\def\cyan{\color{cyan}}
\definecolor{apcolor}{HTML}{b3003b}
\definecolor{afcolor}{HTML}{800080}
\definecolor{lvcolor}{HTML}{DF7401}
\definecolor{mdcolor}{HTML}{01abdf} %mahsa and davide
\definecolor{cbcolor}{HTML}{ff0000}
\definecolor{sccolor}{HTML}{cc5500} %stefano
\definecolor{sgcolor}{HTML}{00cc7a}